\documentclass[twocolumn,american,prl]{revtex4}
\usepackage[T1]{fontenc}
\usepackage[latin1]{inputenc}
\usepackage{graphicx}

\makeatletter

\usepackage{babel}
\makeatother
\begin{document}

\title{Nodal solitons and the nonlinear breaking of discrete symmetry}

\author{Albert Ferrando, Mario Zacarés and Pedro Andrés.}

\affiliation{Departament d'Òptica, Universitat de València. Dr. Moliner, 50. E-46100
Burjassot (València), Spain.}

\author{Pedro Fernández de Córdoba. }

\affiliation{Departamento de Matemática Aplicada, Universidad Politécnica de Valencia.
Camino de Vera, s/n. E-46022 Valencia, Spain.}

\author{Juan A. Monsoriu.}

\affiliation{{\small Departamento de Física Aplicada, Universidad Polit\'{e}cnica
de Valencia. Camino de Vera, s/n. E-46022 Valencia, Spain}.}

\date{\today}

\begin{abstract}
We present a new type of soliton solutions in nonlinear photonic systems
with discrete point-symmetry. These solitons have their origin in
a novel mechanism of breaking of discrete symmetry by the presence
of nonlinearities. These so-called nodal solitons are characterized
by nodal lines determined by the discrete symmetry of the system.
Our physical realization of such a system is a 2D nonlinear photonic
crystal fiber owning $\mathcal{C}_{6v}$ symmetry.
\end{abstract}


\maketitle
Symmetry is one of the most powerful and elegant concepts in physics.
The use of group theory provides an extraordinary mathematical tool
to classify solutions according to the symmetries of the physical
system. In recent years, the increasing interest in physical systems
owning 2D discrete symmetries, such as 2D nonlinear photonic crystals
or Bose-Einstein condensates in 2D periodic potentials, raises the
question of utilizing group theory as an analysis tool. Certainly,
this approach is a standard in solid state physics. Its use in the
topic of photonic crystals is less extended and it has been traditionally
confined to the classification of linear modes \cite{sakoda04}. Its
generalization to the nonlinear case has become of great interest
after the recent experimental observation of fundamental and vortex
solitons in optically-induced 2D nonlinear photonic crystals \cite{fleischer-nature422_147,neshev-prl92_123903,fleischer-prl92_123904}.
In fact, attempts to apply group theory to the analysis of this type
of solutions permitted the analytical prediction of the angular dependence
of vortex solitons in 2D photonic crystals \cite{ferrando-oe12_817}.
Following this approach, in this paper we will use group theory as
a general framework to analyze the role played by nonlinearities in
the realization of discrete symmetry. We will see that discrete symmetry
is realized differently when nonlinearities are present and that a
new phenomenon of discrete-symmetry breaking occurs. The physical
outcome of this general process is the generation of a new type of
solitons with lesser symmetry than that of the original system. 

So that, we start by analyzing the general problem of finding stationary
solutions --$\Phi(x,y,z)=\phi(x,y)\exp i\beta z$-- of a nonlinear
operator of the form:\begin{equation}
\left(L_{0}+L_{NL}(|\Phi|)\right)\Phi=-\frac{\partial^{2}\Phi}{\partial z^{2}},\label{eq:evolution}\end{equation}
where $L_{0}$ is a linear operator (depending on the transverse coordinates
$\mathbf{x}_{t}=(x,y)$ only) invariant under a 2D discrete point-symmetry
group $G$ and $L_{NL}$ is a nonlinear operator depending locally
on the modulus of the $\phi$ field. We are interested then in solving
the following nonlinear eigenvalue problem:\begin{equation}
\left(L_{0}+L_{NL}(|\phi|)\right)\phi=\beta^{2}\phi.\label{eq:eigenvalue-eq}\end{equation}
 Since $L_{0}$ is such that $[L_{0},G]=0$ (i.e., it is invariant
under the action of all the elements of the group $G$: $g^{-1}L_{0}g=L_{0},\,\forall g\in G$),
all its eigenmodes have to lie on finite representations of the discrete
group $G$ \cite{hamermesh64}. What we try to determine next is the
effect of the nonlinear term $L_{NL}$ in the symmetry properties
of Eq.(\ref{eq:eigenvalue-eq}) solutions. 

A solution $\phi_{s}$ of Eq.(\ref{eq:eigenvalue-eq}) has to satisfy
the so-called self-consistency condition, namely, $\phi_{s}$ has
to appear as an eigenmode of the operator generated by itself, $L(\phi_{s})\equiv L_{0}+L_{NL}(|\phi_{s}|)$.
From a symmetry point of view, the self-consistency condition implies
that if $\phi_{s}$ belongs to some representation of a finite group
$G'$, then the entire operator $L(\phi_{s})$ has to be invariant
under the same group, $[L(\phi_{s}),G']=0$ (otherwise, $L(\phi_{s})$
would not contain in its spectrum the representation where $\phi_{s}$lies
on). We call this property the group self-consistency condition. 

The simplest attempt to find solutions of Eq.(\ref{eq:eigenvalue-eq})
satisfying the group self-consistency condition is trying functions
that enjoy the full symmetry of the linear operator; i.e., functions
that are invariant under $G$. Functions belonging to the fundamental
representation of $G$ satisfy this property \cite{hamermesh64}:
$\phi_{\mathrm{fund}}^{g}\equiv g\phi_{\mathrm{fund}}=\phi_{\mathrm{fund}},\,\forall g\in G$.
Group self-consistency is satisfied because $g^{-1}L_{NL}g=L_{NL}(|\phi_{\mathrm{fund}}^{g}|)=L_{NL}(|\phi_{\mathrm{fund}}|),\,\forall g\in G$;i.e.,
$[L_{NL},G]=0$ and, thus, $[L(\phi_{\mathrm{fund}}),G]=0$. Solutions
that satisfy this property are called fundamental solitons and they
have been found in different systems of the type described by Eq.(\ref{eq:eigenvalue-eq}).
A less obvious choice is the selection of functions belonging to higher-order
representations of the same symmetry group $G$ of the linear system.
For 2D point-symmetry groups, these higher-order representations can
be either non-degenerated (one-dimensional) or doubly-degenerated
(two-dimensional) \cite{hamermesh64}. Vortex-antivortex solutions,
appearing always as conjugated pairs ($\phi_{v}$, $\phi_{v}^{*}$),
belong to two-dimensional representations of $G$ \cite{ferrando-oe12_817}.
We note that the modulus of a vortex solution is a group invariant
(this is a general property also fulfilled by the modulus of functions
belonging to one-dimensional representations of $G$). Since $|\phi_{v}^{G}|=|\phi_{v}|$,
then $g^{-1}L_{NL}g=L_{NL}(|\phi_{v}^{g}|)=L_{NL}(|\phi_{v}|),\,\forall g\in G$
and, consequently, $[L_{NL},G]=0$ and $[L(\phi_{v}),G]=0$. It is
apparent that vortex solitons also fulfill the group self-consistency
condition.

\begin{figure}
\includegraphics[%
  scale=0.78]{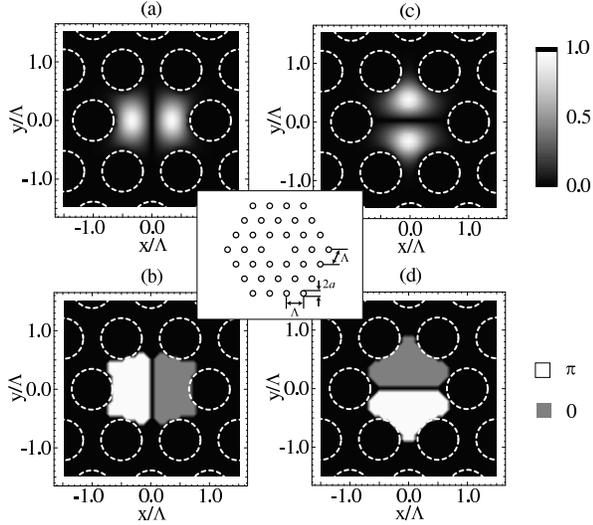}

\caption{Two nodal solitons for $l=1$ ($\Lambda=23\,\mu\mathrm{m}$, $a=8\,\mu\mathrm{m}$,
$\gamma=0.006$ and wavelength $\lambda=1064\,\mathrm{nm}$): (a)-(b)
amplitude and phase, respectively, of the \emph{S} nodal soliton;
(c)-(d) amplitude and phase, respectively, of the \emph{A} nodal soliton.
Inset: schematic transverse representation of a PCF.\label{fig:Amplitudes_and_phases}}
\end{figure}

Up to now, we have considered solutions belonging to representations
of $G$, the symmetry group of the linear operator. For them, the
self-induced nonlinear operator enjoys the same symmetry as its linear
counterpart: $[L_{NL}(|\phi_{s}|),G]=[L_{0},G]=0$. However, we ask
ourselves if solutions with different symmetry than that exhibited
by $G$ can also fulfill the group self-consistency condition. Let
us assume a trial function belonging to a certain representation of
a group $G'$ such that $G'\neq G$. Since the modulus of the function
is $G'$-invariant, $|\phi_{s}^{G'}|=|\phi_{s}|$, the nonlinear operator
is also $G'$-invariant, $[L_{NL}(|\phi_{s}|),G']=0$. If $G'>G$,
the total operator $L(\phi_{s})$ cannot have the $G'$ symmetry because
the linear operator has less symmetry. The linear part of $L(\phi_{s})$
breaks the $G'$ symmetry of the nonlinear part and the group self-consistency
condition cannot be satisfied: $[L(\phi_{s}),G']\neq0$. Thus, we
disregard functions with symmetry higher than $G$. This scenario
changes if one considers functions with lesser symmetry than $G$;
more specifically, functions belonging to representations of a subgroup
$G'$ of $G$ ($G'\subset G$). The difference now with respect to
the previous case is the following: since the linear part is invariant
under the $G$ group, $[L_{0},G]=0$, it is also invariant under any
of its subgroups. Thus, it is also true that $[L_{0},G']=0$. Since,
as before, $|\phi_{s}^{G'}|=|\phi_{s}|$, the nonlinear operator verifies
$[L_{NL}(|\phi_{s}|),G']=0$ and, consequently, the total operator
is $G'$-invariant, $[L(\phi_{s}),G']=0$. The function then fulfills
the group self-consistency condition for the subgroup $G'$. Therefore,
this type of functions can also be solution of the eigenvalue Eq.(\ref{eq:eigenvalue-eq})
(note that solutions with no symmetry also verify the group self-consistency
condition since the identity transformation constitutes a subgroup
of any group $G$). We cannot guarantee that they are indeed solutions
until we solve Eq.(\ref{eq:eigenvalue-eq}) with the constraint $\phi\in D(G')$
($D(G')$ being a representation of $G'$), since the trivial solution
($\phi=0$) is always valid. In this sense, group self-consistency
is a necessary but not sufficient condition. However, if these solutions
are found, they would be characterized by the symmetry breaking pattern
$G\rightarrow G'\subset G$. In other words, we would say that, for
those solutions, nonlinearity breaks the discrete symmetry of the
linear system.

Our specific physical system is a triangular photonic crystal fiber
(PCF), although similar results can be expected in other 2D photonic
crystals. We study the propagation of the electric component of a
monochromatic electromagnetic field (at fixed polarization: $\mathbf{E}=\phi\mathbf{u}$,
$|\mathbf{u}|=1$). PCF's are thin silica fibers possessing a regular
array of holes extending the entire fiber length and characterized
by the hole radius $a$ and the spatial period $\Lambda$ of the photonic
crystal cladding (see inset in Fig. \ref{fig:Amplitudes_and_phases}).
When silica nonlinearity is not neglected, a PCF is a particular case
of a 2D nonlinear photonic crystal with a defect (where guidance occurs).
In this case, $L_{0}=\nabla_{t}^{2}+k_{0}^{2}n_{0}^{2}(x,y)$, where
$\nabla_{t}$ is the transverse gradient operator, $k_{0}$ is the
vacuum wave number, and $n_{0}$ is the refractive-index profile function
($n_{0}=n_{(\mathrm{silica})}$ in silica and $n_{0}=1$ in air).
The nonlinear term is $L_{NL}=k_{0}^{2}\gamma\Delta(x,y)|\phi|^{2}$,
where $\Delta$ is the distribution function of nonlinear material
($\Delta=1$ in silica and $\Delta=0$ in air) and $\gamma$ is a
dimensionless nonlinear coupling constant, $\gamma\equiv3\chi_{(\mathrm{silica})}^{(3)}P/(2\varepsilon_{0}cn_{(\mathrm{silica})}A_{0})$
($P$ is the total power and $A_{0}$ is an area parameter: $A_{0}=\pi(\Lambda/2)^{2}$). 

The symmetry group of a triangular PCF is $\mathcal{C}_{6v}$, i.e.,
$[L_{0},\mathcal{C}_{6v}]=0$. This group is constituted by discrete
$\pi/3$-rotations ($R_{\pi/3}$) plus specular reflections with respect
to the $x$ and $y$ axes: $\theta\stackrel{R_{x}}{\rightarrow}-\theta$
and $\theta\stackrel{R_{y}}{\rightarrow}\pi-\theta$, in polar coordinates.
Solutions with the PCF $\mathcal{C}_{6v}$-symmetry have been previously
found in the form of fundamental and vortex solitons \cite{ferrando-oe11_452,ferrando-oe12_817}.
We will focus now on new solutions belonging to the subgroup $\mathcal{C}_{2v}$
of $\mathcal{C}_{6v}$, formed by $R_{\pi},R_{x}$ and $R_{y}$. Thus,
we study the particular symmetry breaking pattern $G=\mathcal{C}_{6v}\rightarrow G'=\mathcal{C}_{2v}$.
We can explicitly construct functions belonging to the four non-degenerated
representations of $\mathcal{C}_{2v}$ out of functions in the two-dimensional
representations of $\mathcal{C}_{6v}$. The latter functions come
in conjugated pairs ($\phi_{l}$, $\phi_{l}^{*}$) ($l=1,2)$, whose
angular dependence is fixed by symmetry: $\phi_{l}=r^{l}e^{il\theta}\phi_{l}^{s}(r,\theta)\exp[i\phi_{l}^{p}(r,\theta)]$,
where $\phi^{s}(r,\theta)$ is a scalar function, characterized by
$\phi^{s}(r,\theta+\pi/3)=\phi^{s}(r,\theta)$ and $\phi^{s}(r,-\theta)=\phi^{s}(r,\pi-\theta)=\phi^{s}(r,\theta)$,
and $\phi^{p}(r,\theta)$ is a pseudoescalar function characterized
by $\phi^{p}(r,\theta+\pi/3)=\phi^{p}(r,\theta)$ and $\phi^{p}(r,-\theta)=\phi^{p}(r,\pi-\theta)=-\phi^{p}(r,\theta)$.
Let us consider the two following linear combinations ($l=1,2$):
$1/\sqrt{2}\left[\phi_{l}\pm\phi_{l}^{*}\right]$). By writing the
angular dependence of $\phi_{l}$, the new functions adopt the form:\begin{equation}
\phi_{\delta}^{l}(r,\theta)=\sqrt{2}r\phi_{l}^{s}(r,\theta)\cos\left[l\theta+\phi_{l}^{p}(r,\theta)+\delta\right],\, l=1,2.\label{eq:nodal-soliton}\end{equation}
where $\delta=0,\pi/2$. The four different type of solutions given
by Eq.(\ref{eq:nodal-soliton}) belong to the four different one-dimensional
representations of the $\mathcal{C}_{2v}$ group; i.e., in Hamermesh's
notation: $\phi_{0}^{1}\in B_{1}$,$\phi_{\pi/2}^{1}\in B_{2}$,$\phi_{0}^{2}\in A_{1}$,$\phi_{\pi/2}^{2}\in A_{2}$.
According to our previous general argument, the $\phi_{\delta}^{l}$
functions can be solutions of Eq.(\ref{eq:eigenvalue-eq}) for a PCF.
Therefore, we solve Eq.(\ref{eq:eigenvalue-eq}) with the ansatz given
by Eq.(\ref{eq:nodal-soliton}) by means of the Fourier iterative
method previously used in Refs.\cite{ferrando-oe11_452} and \cite{ferrando-oe12_817}
to find fundamental and vortex soliton solutions in PCF's. This method
ensures that the group representation is preserved in the iterative
process. Starting from a seed function of the form (\ref{eq:nodal-soliton}),
the method either finds the trivial solution or converges to a solution
belonging to one of the representations of $\mathcal{C}_{2v}$.

Solutions of Eq.(\ref{eq:eigenvalue-eq}) of the form given by Eq.(\ref{eq:nodal-soliton})
are indeed found. They are characterized by nodal lines determined
by symmetry through the implicit equation $\cos\left[l\theta+\phi_{l}^{p}(r,\theta)+\delta\right]=0$
($l=1,2$). For this reason, we call them nodal solitons. The solution
with $\delta=0$ corresponds to the symmetric (\emph{S}) $\mathcal{C}_{6v}$
combination $\phi_{S}^{l}\equiv1/\sqrt{2}(\phi_{l}+\phi_{l}^{*})$
and that with $\delta=\pi/2$ to the antisymmetric (\emph{A}) one
$\phi_{A}^{l}\equiv i/\sqrt{2}(\phi_{l}-\phi_{l}^{*})$ ($l=1,2$).
Note that the $\phi_{l}$ function is not a solution of the Eq.(\ref{eq:eigenvalue-eq})
because the superposition principle does not hold. This function can
only be approximated by a vortex solution in the linear regime ($\gamma\approx0$).
Although the complete structure of nodal lines could be rather intrincate,
\emph{S} and \emph{A} nodal solitons are characterized by principal
nodal lines: a single principal line for $l=1$ solitons and two orthogonal
principal lines for $l=2$. In our simulations, we have found this
four different types of nodal solitons. Nevertheless, we will show
here results corresponding to the \emph{S} and \emph{A} nodal soliton
solutions with $l=1$ only. In Fig. \ref{fig:Amplitudes_and_phases}
we show the amplitude and phase of \emph{S} and A nodal solitons,
respectively. As predicted by the nodal line condition, the $l=1$,
\emph{S} nodal soliton presents a single vertical nodal line, whereas
for the \emph{A} soliton this line is horizontal.

\begin{figure}
\includegraphics{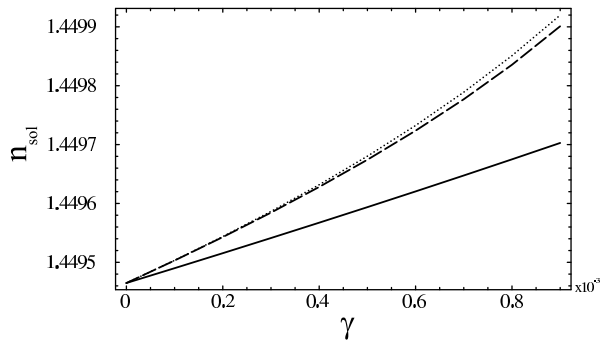}

\caption{Effective index of a soliton solution, $n_{\mathrm{sol}}$ vs the
nonlinear coupling $\gamma$ for symmetric (dotted line) and antisymmetric
(dashed line) solitons and vortex and antivortex solitons with $l=1$
(solid line).\label{fig:Effective-index}}
\end{figure}

The role played by symmetries can be clearly envisaged by analyzing
the diagram $n_{\mathrm{sol}}$$vs.$ $\gamma$, where $n_{\mathrm{sol}}=\beta/k_{0}$
for different soliton solutions. In Fig. \ref{fig:Effective-index}
we represent the curves corresponding to the \emph{S} and \emph{A}
nodal solitons as well as the curves for vortex solitons reported
in Ref.\cite{ferrando-oe12_817}. In the linear case ($\gamma=0$),
the superposition principle holds and, therefore, the symmetric and
antisymmetric combinations of the linear modes $\phi_{1}$ and $\phi_{1}^{*}$
(the linear modes of vortex-type with $l=1$) are degenerated solutions.
In the linear case, all of them (\emph{S}, \emph{A}, $\phi_{1}$ and
$\phi_{1}^{*}$ modes) belong to the same representation of the $\mathcal{C}_{6v}$
group ($l=1$, or $E_{2}$ in Hamermesh's notation) and, for this
reason, they all have the same effective index. The presence of the
nonlinearity changes this scenario. It provides different options
for the realization of the discrete symmetry. Vortex solitons realize
the discrete symmetry of the linear system, $[L(|\phi_{1}|),\mathcal{C}_{6v}]=0$,
whereas nodal solitons break this symmetry into its $\mathcal{C}_{2v}$
subgroup, $[L(|\phi_{S,A}|),\mathcal{C}_{2v}]=0$. Consequently, their
corresponding eigenvalues are different since they are no longer related
by the original symmetry that they enjoyed in the linear case ($\gamma=0$).
Moreover, group theory predicts that the vortex and antivortex solutions
must have the same effective index as they provide the same total
operator $L(|\phi_{1}|)$ and they belong to the same representation
($E_{2}$) of it. This is not the case for nodal solitons. $\phi_{S}$
and $\phi_{A}$ are not in the same representation of $\mathcal{C}_{2v}$.
This fact immediately implies that its corresponding eigenvalues must
be different. Curves in Fig. \ref{fig:Effective-index} explicitly
manifest this feature. For small nonlinearities, all curves appear
as nearly-degenerated, but as the value of $\gamma$ increases, a
growing gap between nodal solitons and vortices occurs. Even larger
values of $\gamma$ permit to manifest the existing gap between the
\emph{S} and \emph{A} nodal solitons. 

\begin{figure}
\includegraphics{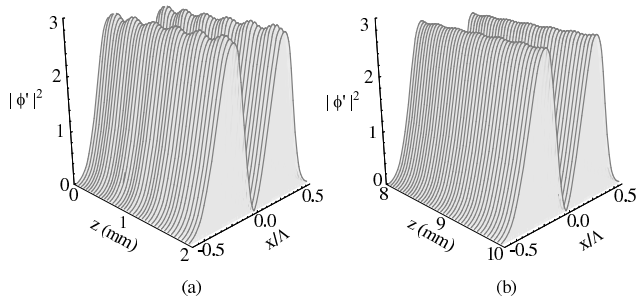}

\caption{Evolution of a diagonal perturbation of a \emph{S} nodal soliton
($|\phi'(x,0,z)|^{2}$) showing asymptotic stability.\label{fig:Evolution-diagonal}}
\end{figure}

\begin{figure}
\includegraphics[%
  scale=0.78]{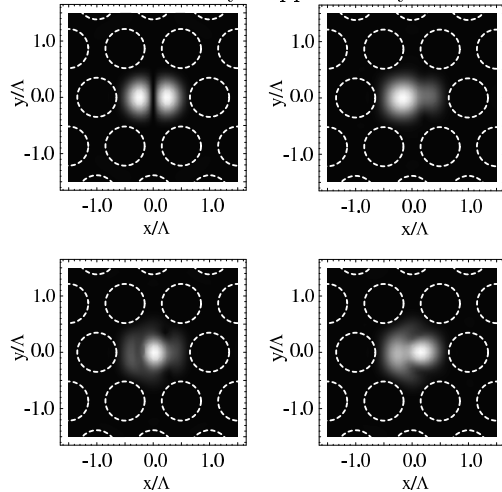}

\caption{Four snap-shots of the evolution of a non-diagonal perturbation of
a \emph{S} nodal soliton. In this case, an oscillatory instability
occurs.\label{fig:non-diagonal_perturbation}}
\end{figure}

In order to check the stability of nodal solitons we need to solve
the evolution Eq. (\ref{eq:evolution}) after perturbing the solution:
$\phi'_{S,A}=\phi_{S,A}+\delta\phi$. The stability analysis follows
that of vortex solitons in Ref. \cite{ferrando-oe12_817}. In there
we introduced the concepts of diagonal and non-diagonal perturbations.
For one-dimensional representations, a diagonal perturbation is defined
as that that preserves the representation in which the solution lies
on. In the present case ---in which $\phi_{S}$ and $\phi_{A}$ belongs
to the $B_{1}$ and $B_{2}$ one-dimensional representations of $\mathcal{C}_{2v}$,
respectively---, this definition implies that $\phi'_{S}\in B_{1}$
and $\phi'_{A}\in B_{2}$. Explicit examples of such perturbations
are scaled solutions: $\phi'_{S,A}=(1+\epsilon)\phi_{S,A}$, $\epsilon\neq0$.
Evolution yields numerical evidence that nodal solitons are stable
under such perturbations, as shown in Fig \ref{fig:Evolution-diagonal}
in which a diagonal perturbation (scaled solution) is applied. However,
non-diagonal perturbations, taking the perturbed solution out of its
original representation, provide instabilities. These instabilities
are of the oscillatory type, as shown in Fig. \ref{fig:non-diagonal_perturbation},
and they can be understood as a simultaneous oscillation among modes
belonging to all the representations of $\mathcal{C}_{2v}$. This
instability pattern, however, shows no trace of pseudo-soliton collapse
nor of transverse ejection of pseudo-solitons typical of Kerr nonlinearities
in homogeneous media treated in the paraxial approximation. This particular
behavior was first observed in vortices in PCF's \cite{ferrando-oe12_817}.
Since the self-focussing instability seems to be rooted in the paraxial
approximation \cite{akhmediev-ol18_411}, a plausible explanation
of its absence is the non-paraxial nature of evolution in this case.
Absence of ejection can be qualitative understood by the inhibition
of transverse radiation induced by the photonic crystal cladding.

An interesting interpretation of nodal solitons is as interacting
pseudo-solitons. It can be proven that a \emph{S} nodal soliton can
be written as $\phi_{S}=\phi_{0}(x+x_{0},y)-\phi_{0}(x-x_{0},y)$,
$\phi_{0}$ being a localized function in the fundamental representation
of $\mathcal{C}_{6v}$. In the case that $\phi_{0}$ is a sufficiently
localized function (large nonlinear coupling $\gamma$ or strong lattice
index contrast), $\phi_{0}$ can be approximated by a fundamental
soliton solution. Then, a nodal soliton can be envisaged as a pair
of weakly interacting pseudo-solitons. Like in an homogeneous medium
this interaction is repulsive \cite{stegeman-science286_1518}. In
our case, nor $\gamma$ nor the index contrast are necessarely large,
consequently, the soliton-soliton interaction cannot longer be considered
weak. However, the group theory antisymmetric decomposition in terms
of localized solutions in the fundamental representation of $\mathcal{C}_{6v}$
remains valid. In this way, the concept of nodal soliton generalizes
the idea of interacting pseudo-solitons into a regime of strong particle
coupling (intermediate $\gamma$ and index contrast). In the weak
soliton-interaction regime, new soliton solutions have been recently
found that can also be interpreted in the context of group theory
reported in this paper \cite{alexander-prl93_63901}.

We are thankful to H. Michinel for useful discussions. This work was
financially supported by the Plan Nacional I+D+I (grant TIC2002-04527-C02-02),
Ministerio de Ciencia y Tecnolog\'{\i}a (Spain) and FEDER funds.
Authors also acknowledge the financial support from the Generalitat
Valenciana, Spain (grants Grupos03/227 and GV04B-390). M. Z. gratefully
acknowledges Fundaci\'{o}n Ram\'{o}n Areces grant.

\bibliographystyle{/usr/share/texmf/bibtex/bst/revtex4/apsrev}
\bibliography{bib_general}

\end{document}